\documentclass{IEEEtran}
\usepackage{cite}
\usepackage{amsmath,amssymb,amsfonts}
\usepackage{graphicx}
\usepackage{textcomp,nicefrac}
\usepackage[switch]{lineno}
\usepackage{xcolor}

\def\BibTeX{{\rm B\kern-.05em{\sc i\kern-.025em b}\kern-.08em
T\kern-.1667em\lower.7ex\hbox{E}\kern-.125emX}}
\begin{document}
\title{Temperature-Dependent Performance of NaI(Tl) Crystal with Dual-Channel SiPM Readout for Low-Mass Dark Matter Searches}
\author{W.~K.~Kim, H.~Y.~Lee, K.~W.~Kim, and H.~S.~Lee
\thanks{This work was supported by the Institute for Basic Science (IBS),
Republic of Korea under project code IBS-R016-A1.}
\thanks{W.~K.~Kim and K.~W.~Kim are with the Center for Underground Physics, Institute for Basic Science (IBS), Daejeon 34126, Republic of Korea.}
\thanks{H.~Y.~Lee is with the Center for Exotic Nuclear Studies, Institute for Basic Science (IBS), Daejeon 34126, Republic of Korea.}
\thanks{H.~S.~Lee is with the Center for Underground Physics, Institute for Basic Science (IBS), Daejeon 34126, Republic of Korea and IBS School, University of Science and Technology (UST), Daejeon 34113, Republic of Korea.}
\thanks{Corresponding author: W.~K.~Kim (e-mail: wonkyung@ibs.re.kr) and K.~W.~Kim (e-mail: kwkim@ibs.re.kr)}
}

\maketitle

\begin{abstract}
We report the first temperature-dependent characterization of a NaI(Tl) crystal readout by two silicon photomultipliers (SiPMs) directly coupled to opposite ends of the crystal for rare-event searches. A $6 \text{ mm} \times 6 \text{ mm} \times 13 \text{ mm}$ NaI(Tl) crystal was directly coupled to two SiPMs and characterized in a liquid nitrogen-cooled cryostat over a temperature range of 94$-$293\,K. The light yield, energy resolution, and scintillation decay time were measured using $\gamma$-ray peak from a $^{241}$Am source. After correcting for optical crosstalk contributions, the light yield increased, reaching $17.7 \pm 1.1$ photoelectrons/keV at 238\,K, corresponding to a 34.5\% enhancement relative to room temperature (293\,K). Furthermore, dual-channel configuration effectively suppresses random thermal noise via coincidence triggers, which together with the observed increase in light yield, provides a critical pathway toward lowering the energy threshold for dark matter and coherent elastic neutrino$-$nucleus scattering searches.
\end{abstract}


\section{Introduction}
\label{sec:introduction}
Thallium doped sodium iodide, NaI(Tl), is one of the most widely used scintillating crystals in direct searches for dark matter (DM)~\cite{adhikari2018,antonello2019,amare2019b,Bernabei:2020mon} and coherent elastic neutrino$-$nucleus scattering (CE$\nu$Ns)~\cite{konovalov2024}. The COSINE-100 experiment, which employs low-background NaI(Tl) crystals coupled to photomultiplier tubes (PMTs)~\cite{adhikari2018b, lee2025}, was designed to test the DAMA/LIBRA annual modulation claim with the same target material and has reported constraints in tension with that result~\cite{carlin2025}. Extending the reach of NaI(Tl) detectors to the sub-MeV mass DM region, requires low-energy thresholds, since the corresponding nuclear recoil energies fall well below the keV scale~\cite{goodman1985a,schumann2019,Billard:2021uyg}. For scintillation detectors the achievable threshold is governed primarily by the light yield, thus increasing the number of detected photoelectrons (PEs) is the most direct route toward lowering the energy threshold and thereby probing the lighter weakly interacting massive particles (WIMPs).
\begin{figure}[t]
\centerline{\includegraphics[width=3.5in]{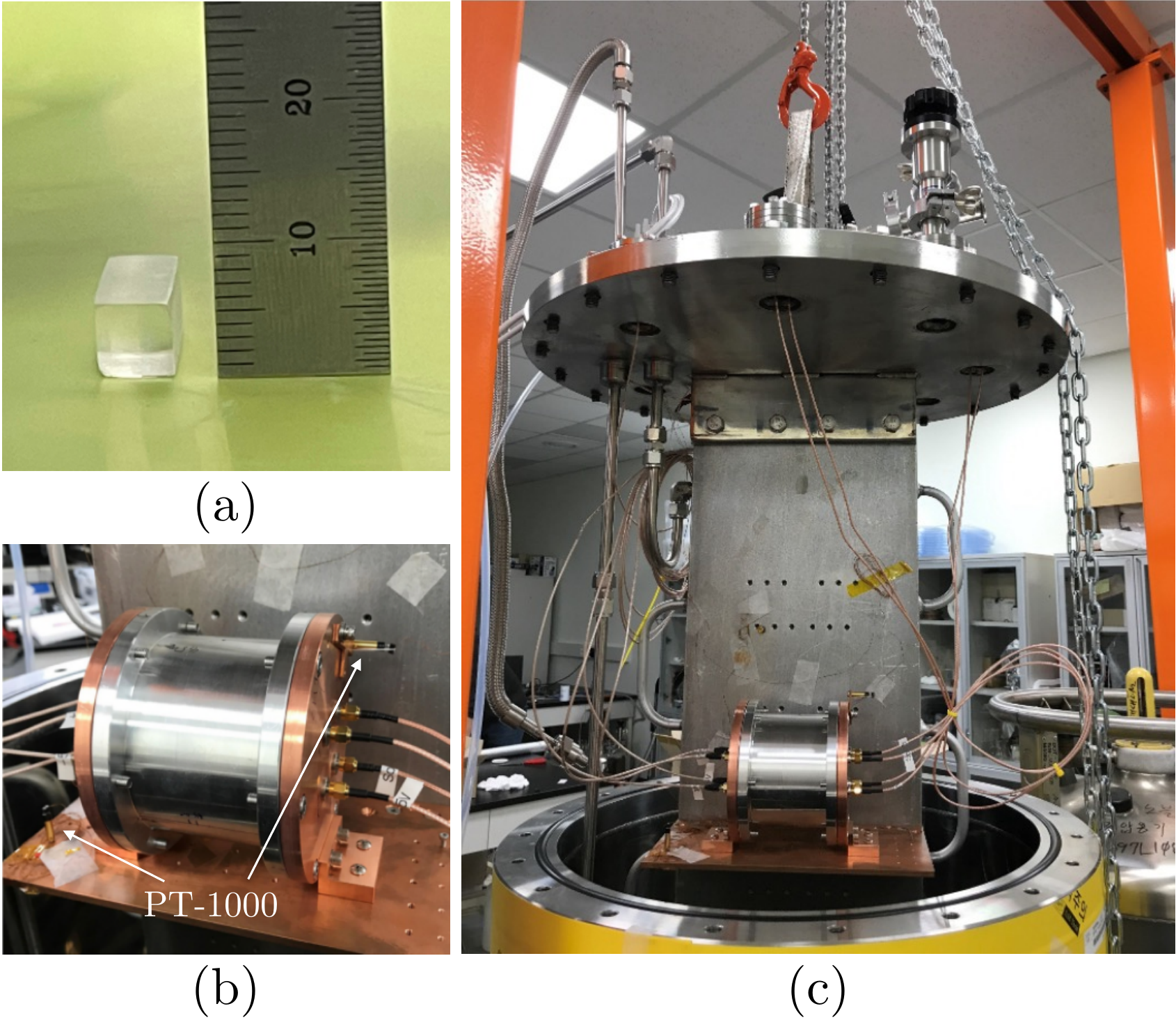}}
\caption{(a) Dimensions of $6 \text{ mm} \times 6 \text{ mm} \times 13 \text{ mm}$ NaI(Tl) crystal. (b) Encapsulation. (c) Cryostat vacuum chamber.}
\label{setup}
\end{figure}
\begin{figure}[t]
\centerline{\includegraphics[width=3.5in]{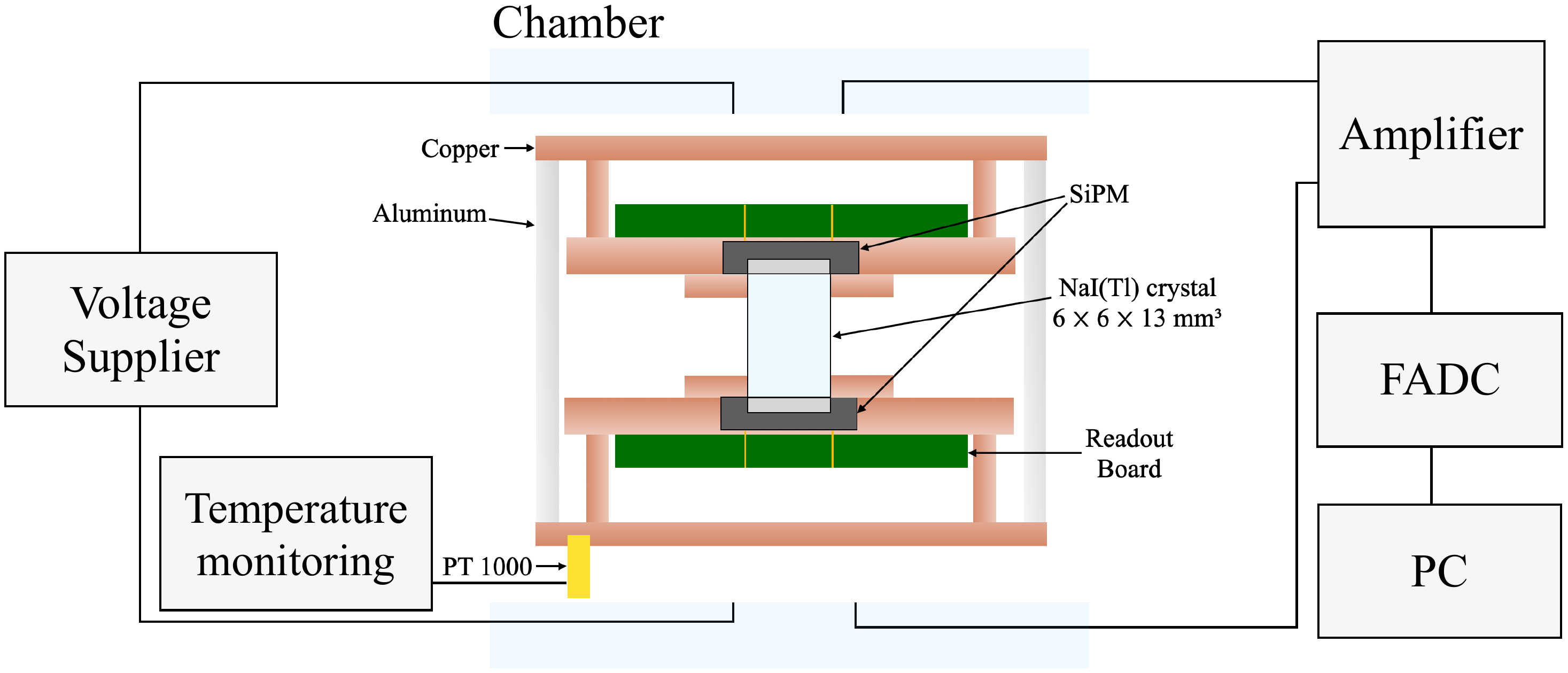}}
\caption{Schematic diagram of the dual-channel SiPM$-$NaI(Tl) experimental setup.}
\label{setup_schematic}
\end{figure}
Two complementary handles are available for improving the NaI(Tl) detector performance: exploiting the temperature dependence of the scintillation light yield, and replacing the photosensor itself. The scintillation light yield of NaI(Tl) is reported to increase at lower temperatures~\cite{sailer2012,lee2022a,park2026, martinenghi2025}. Conventional NaI(Tl) detectors, however, operate at room temperature with PMTs attached to the crystal surfaces to read out the scintillation signals. PMTs are bulky, have dark current, and contain a non-negligible amount of radioactive contaminants, which introduce low-energy backgrounds that raise the analysis threshold. Nonetheless, the current COSINE-100 experiment has lowered the energy threshold to 0.7\,keV by leveraging various methods of event selection~\cite{adhikari2018b, yu2024}. PMTs could be replaced by SiPMs, which offer high photon detection efficiency (PDE), high gain, and lower intrinsic radioactivity~\cite{otte2017}. Their main drawback for low-energy applications is a high dark count rate (DCR), which is strongly suppressed at low-temperature~\cite{kim2025, lee2022, DINU2015275, ANFIMOV2021165162}. Notably, both handles point toward the same direction—low-temperature operation—motivating a systematic study of NaI(Tl) crystal performance coupled to two SiPM channels over the temperature range 94$-$293\,K, carried out as part of COSINE's SiPM R\&D campaign~\cite{kim2025, lee2022}. While the previous study~\cite{lee2022} used a single SiPM channel, the present work extends the readout to two SiPMs coupled to opposite ends of the crystal, following the PMT dual-channel of the COSINE detectors~\cite{adhikari2018b}. The multi-channel increases the total number of detected PEs and enables coincide between the channels to suppress noise triggering. This configuration is a step toward full SiPM coverage of all crystal surfaces.

\section{Experimental setup}
\label{sec:setup}

A NaI(Tl) crystal with dimensions of $6 \text{ mm} \times 6 \text{ mm} \times 13 \text{ mm}$ was used for the measurement as shown in Fig.~\ref{setup}(a). It was cut from an ingot grown in the low-background NaI(Tl) development program of the COSINE-200, reported as NaI-036 in~\cite{COSINE:2020egt}. Several layers of soft polytetrafluoroethylene (PTFE) sheet were wrapped around the exposed crystal surfaces to reflect scintillation photons. Two Hamamatsu S13360-6050CS SiPMs with quartz window type were directly attached to the two end faces of the crystal to maximize the light collection efficiency. To prevent from the hygroscopic degradation, the detector was encapsulated with package made with stainless steel and copper as shown in Fig.~\ref{setup}(b). 

The detector was installed in a cryostat vacuum chamber whose cooling was achieved by circulating liquid nitrogen (LN$_2$) through a stainless steel tube, as in Fig.~\ref{setup}(c). Upon LN$_2$ injection, the temperature dropped to 94\,K within 5\,h and then increased gradually back toward room temperature (293\,K). The lowest temperature of 94\,K was limited by the indirect cooling of chamber design and by the heat dissipated through the pre-amplifiers on the readout boards.
Two temperature sensors of PT-1000 were mounted on the outer copper case of the encapsulation and the copper plate where the detector was installed and connected to a Lake Shore 336 for monitoring. Data were taken every 10\,K during both the cooling and warming phases over two cooling cycles; only the warming phase data were adopted in the analysis to reduce the systematic uncertainty associated with the temperature stabilization. A residual systematic uncertainty of $\pm1$\,K on the reported temperatures is assigned from the operation principle of this chamber.
Two SiPMs were biased through readout boards providing pre-amplification, bias voltage supply and signal readout, as illustrated in Fig.~\ref{setup_schematic}. Two KEITHLEY 2635B SourceMeters supplied the SiPM bias voltages, while a KEYSIGHT E3630A voltage supply powered the boards. The board signals were further amplified by a factor of 30 with an external amplifier and digitized by a 125\,MHz, 12-bit flash-analog-to-digital converter (FADC). To accommodate both the long phosphorescence of NaI(Tl) and the SiPM after-pulses, a 32\,$\mu$s waveform was recorded for each triggered event. 

\begin{figure}[t]
\centerline{\includegraphics[width=3.5in]{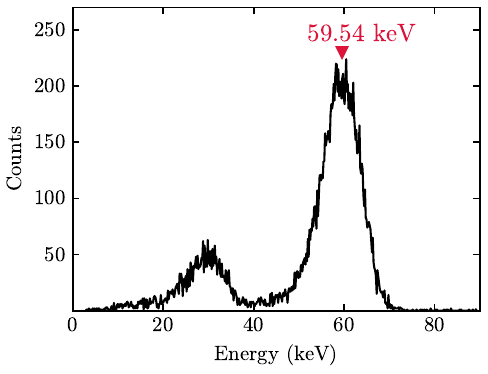}}
\caption{A representative $^{241}$Am energy spectrum measured with the dual-channel NaI(Tl)--SiPM detector at 238\,K. The 59.54\,keV $\gamma$-ray peak was used to calibrate the detector response. The energy resolution at this peak is $\sigma$/m = $5.3 \pm 0.2$ \% (FWHM/E = $12.5 \pm 0.5$ \%).}
\label{fig:ampeak}
\end{figure}

The detector calibration was performed with the 59.54\,keV $\gamma$-ray peak from a $^{241}$Am source, which was placed at the center of the encapsulation. Fig.~\ref{fig:ampeak} shows a representative energy spectrum obtained with the $^{241}$Am calibration source at 238\,K.
Because the SiPM gain is affected by the temperature~\cite{lee2022, DINU2015275, ANFIMOV2021165162}, all temperature-dependent measurements were performed at a fixed gain, so that the height of single-PE remained comparable across temperatures.

\section{Performance of NaI(Tl)--SiPM detector from 94 to 293\,K}
\label{sec:measurements}

\subsection{SiPM characterization}
\label{subsec:sipm}

\begin{figure}[htb]
\centerline{\includegraphics[width=3.5in]{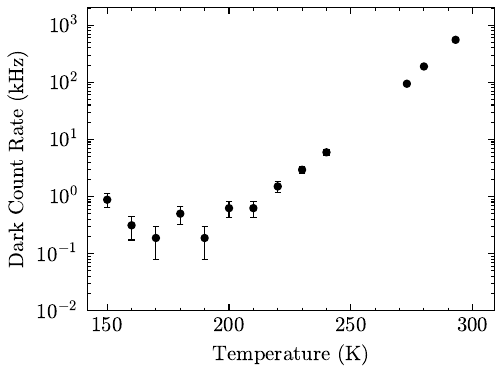}}
\caption{
    Temperature dependence of the DCR of the SiPM used in this work, adapted from~\cite{kim2025} for the same SiPM model. The DCR decreased by more than three orders of magnitude between room temperature and 150\,K. 
    }
\label{fig:dark}
\end{figure}

Two noise components dominate the SiPM response relevant to a low-energy scintillation measurement: the DCR and the optical crosstalk probability.
As shown in Fig.~\ref{fig:dark}, the DCR decreases dramatically at cryogenic temperatures because of the reduced thermal generation probability~\cite{otte2017}.
Crosstalk is categorized into internal and external terms, depending on where the secondary avalanche happens. Internal crosstalk arises when photons emitted during an avalanche initiate a secondary avalanche in a nearby pixel of the same SiPM. External crosstalk, on the other hand, occurs when these photons escape from one SiPM and trigger an avalanche in another SiPM located nearby~\cite{kim2025,Wang:2022ekc, guan}. In the present setup, the two SiPMs were attached to opposite ends of the transparent NaI(Tl) crystal; therefore, photon escaping from one SiPM could propagate through the crystal and reach the other SiPM. The external crosstalk contribution was therefore also considered in the analysis correction. 
Following in method in~\cite{lee2022,kim2025}, the internal crosstalk probability was obtained as the ratio of the number of charge-weighted multi-PEs events to that of single-PE events. The external crosstalk probability was taken from our previous measurement with the same model and type of SiPM coupled to both end faces of a PTFE wrapped crystal~\cite{kim2025}, in which the bias voltage of one SiPM was kept fixed while the facing SiPM was turned off, turned on, and operated at overvoltages increased in 1\,V steps. The external crosstalk probability was derived from the resulting increase in the crosstalk probability of the fixed bias voltage SiPM. At an overvoltage of 3\,V, the calculated internal crosstalk and external crosstalk probabilities are $19.3 \pm 3.2\%$ and $7.6 \pm 0.4\%$, respectively. Because the values were determined with not only the SiPM but also coupled to the crystal and the PTFE reflector, they include photons reflected back into the SiPMs and are expected to be larger than those of SiPM only measurements, providing conservative correction of the light yield. Since the crosstalk probabilities are reported independent from the temperature~\cite{lightfoot}, the values measured at 293\,K were applied to all temperature data.

\begin{figure}[htb]
\centerline{\includegraphics[width=3.5in]{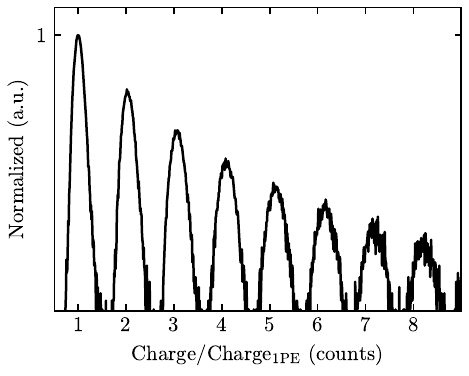}}
\caption{The dark count charge spectrum measured at room temperature is shown. Well-defined peaks corresponding to integer PEs are observed, with the most prominent peak at single-PE. Peaks at multi-PEs are also visible, arising from correlated avalanches induced by optical crosstalks.}
\label{fig:pe}
\end{figure}

\subsection{Light yield and energy resolution}
\label{subsec:energyresolution}
\begin{figure}[htb]
\centerline{\includegraphics[width=3.5in]{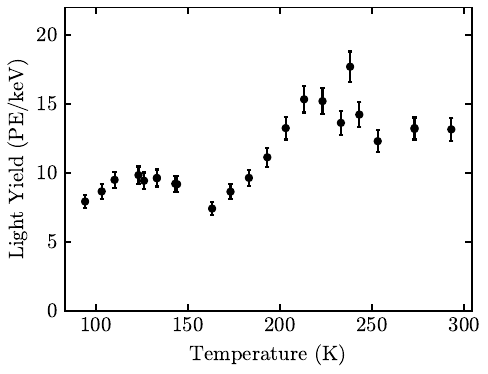}}
\caption{Absolute light yield of the dual-channel SiPM--NaI(Tl) detector as a function of temperature. The maximum of $17.7\pm1.1$\,PE/keV is reached at 238\,K, a 34.5\% increase relative to the room-temperature value of $13.2\pm0.8$\,PE/keV.}
\label{fig:ly}
\end{figure}

The light yield, defined in this paper as the number of detected PEs per unit deposited energy (PE/keV), is shown in Fig.~\ref{fig:ly} as a function of temperature. The behavior agrees with previous reports on cryogenic NaI(Tl) measurement~\cite{sailer2012, lee2022}, where the enhancement of light yield at intermediate temperatures has been attributed to more efficient exciton transfer to Tl luminescence centers together with reduced non-radiative losses~\cite{robert1949}. The light yield was obtained from the charge integrated over the entire 32\,$\mu$s waveform.
A light yield of $13.2\pm0.8$\,PE/keV was obtained at room temperature, rising to a maximum of $17.7\pm1.1$\,PE/keV at 238\,K, a 34.5\% enhancement.

\begin{figure}[htb]
\centerline{\includegraphics[width=3.5in]{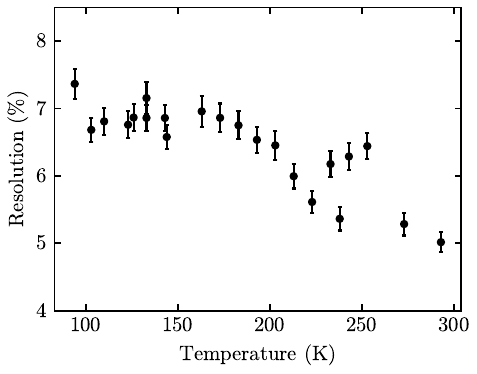}}
\caption{Energy resolution at 59.54\,keV peak as a function of temperature. The best resolution of $5.3 \pm 0.2$\% is obtained at 238\,K.}
\label{fig:resol}
\end{figure}

The energy resolution, defined as the root-mean-square width over the mean ($\sigma/m$) of a single Gaussian fit to the 59.54\,keV peak, is shown in Fig.~\ref{fig:resol}. 
Over most of the measured temperature range, the resolution generally improves with increasing light yield and reaches $5.3\pm0.2$\% at 238\,K. A slightly better resolution is observed at room temperature despite the lower light yield, indicating the resolution is not governed solely by photoelectron statistics. This suggests additional non-statistical contributions. Possible sources include residual gain variations and temperature-dependent changes in light collection uniformity, potentially arising from differential thermal contraction of the detector components.

\subsection{Decay time}
\label{subsec:decaytime}
\begin{figure}[t]
\centerline{\includegraphics[width=3.5in]{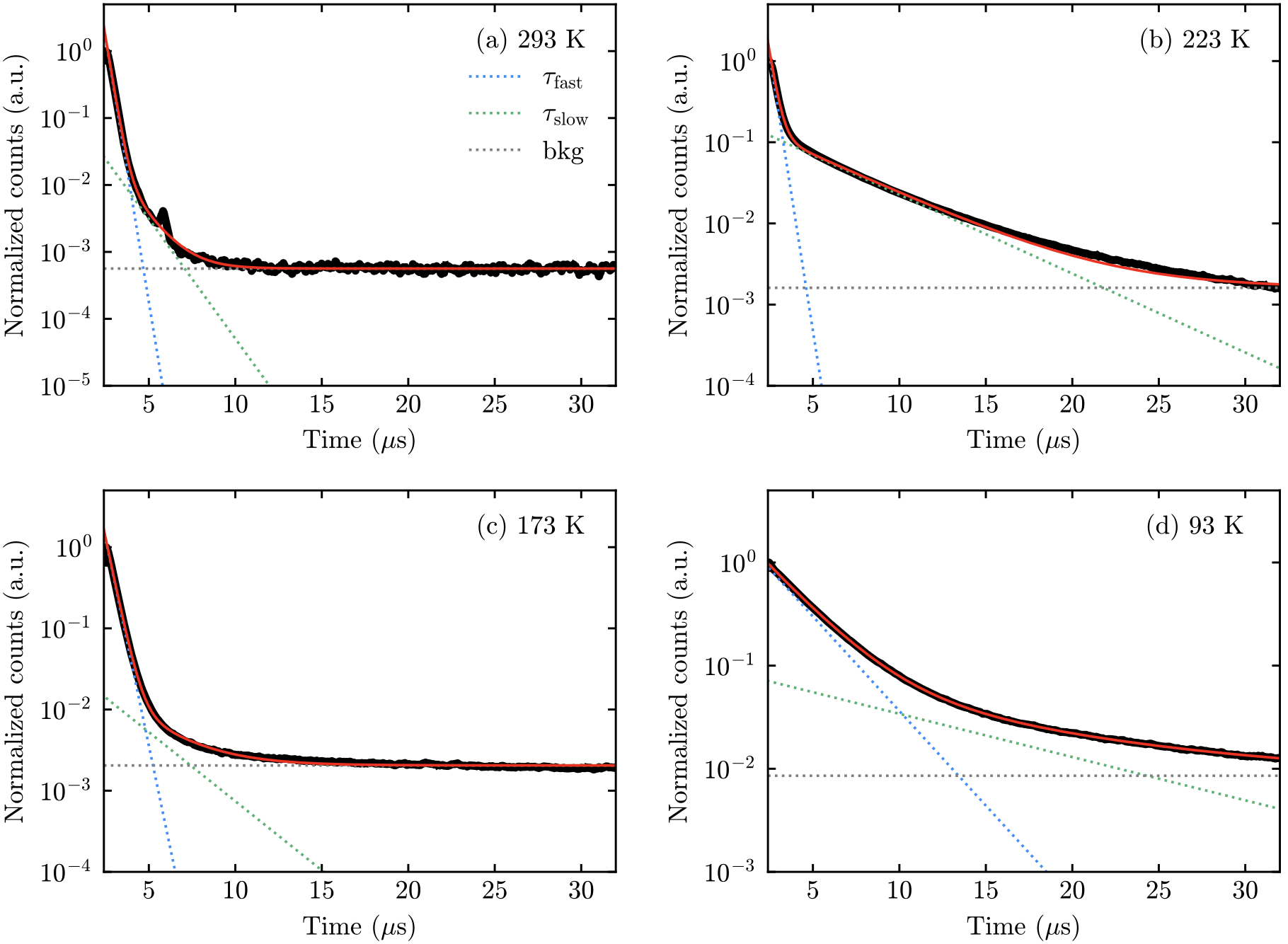}}
\caption{Accumulated scintillation waveforms of SiPM--NaI(Tl) measured at different temperatures: (a) 293\,K, (b) 223\,K, (c) 173\,K, and (d) 93\,K. The solid red line represents two-exponential fits to the decay component, from which the decay time constants are extracted.}
\label{fig:decayfit}
\end{figure}

\begin{figure}[t]
\centerline{\includegraphics[width=3.5in]{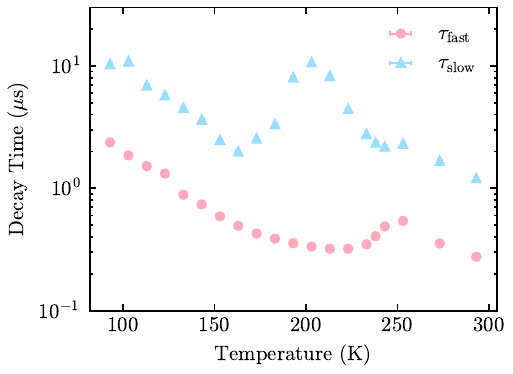}}
\caption{Fast ($\tau_{fast}$) and slow ($\tau_{slow}$) decay constants of the NaI(Tl) crystal as a function of temperature, extracted from the two exponential components fit of Eq.~\eqref{eq:decay}. The slow component follows the trend of the light yield, with its largest contribution near 230\,K.}
\label{fig:decaytime}
\end{figure}

The scintillation decay time of NaI(Tl) crystal with SiPM was studied by accumulating scintillation events at each temperature and fitting them with a two-component exponential including a flat background component,
\begin{equation}
    F(t) = a_1 \exp{[\frac{-(t-t_0)}{\tau_{fast}}]} + a_2 \exp{[\frac{-(t-t_0)}{\tau_{slow}}] + const.},
    \label{eq:decay}
\end{equation}
where $\tau_{fast}$ and $\tau_{slow}$ are the fast and slow decay constants, $a_1$ and $a_2$ are normalization factors, and $t_0$ is the rising edge position. At the room temperature, the constant term mainly dominated by SiPM dark counts, while at low-temperatures, it mainly arises from components with time scales much longer than the time window, such as SiPM after-pulses and the phosphorescence of NaI(Tl).
The fitted waveforms are presented in Fig.~\ref{fig:decayfit}.
The fast component is attributed to the prompt capture of self-trapped excitons at the Tl activator centers, whereas the slow component arises from delayed recombination in which excitons or charge carriers migrate through the lattice via thermally assisted hopping and diffusion, before reaching the activator centers~\cite{Dietrich1973}. The fitted decay constants as a function of temperature are shown in Fig.~\ref{fig:decaytime}. The trend of the decay constant at low-temperatures is consistent with previous measurements~\cite {lee2022, sailer2012}. The slow component reaches its largest relative contribution near 230\,K, and a reversal of the fast and slow fractions near 240\,K is observed, consistent with PMT measurements~\cite{sailer2012}.


\section{Event selection and detector threshold}
\label{sec:threshold}
\begin{figure}[t]
\centerline{\includegraphics[width=3.5in]{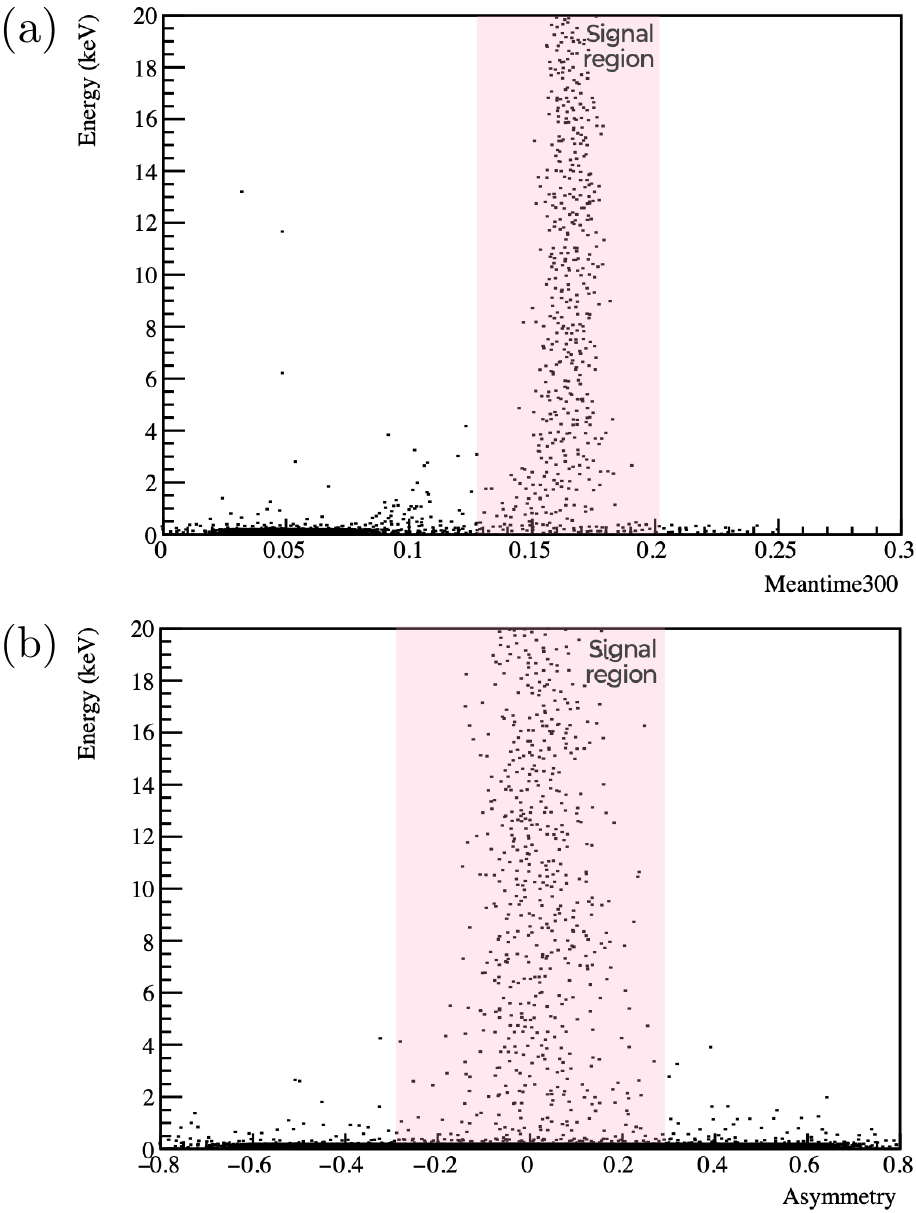}}
\caption{Two-dimensional distributions of energy versus (a) \textit{meantime} parameter and (b) detector charge asymmetry obtained  from  15.5\,h background run at 240\,K under the dual-channel detector configuration. The pink shaded regions indicate the signal region for the event selection cut.}
\label{fig:es}
\end{figure}

In addition to improving the light yield and energy resolution, the dual-channel configuration offers a further advantage for lowering the analysis threshold; it enables noise rejection through coincidence and correlation between the two channels. Unlike a single-channel detector, this readout provides additional information through the detector asymmetry. We define two observables from the waveforms.
First, the detector charge asymmetry, 
\begin{equation}
    \text{Asymmetry}\equiv \frac{q_\text{CH1}-q_\text{CH2}}{q_\text{CH1}+q_\text{CH2}},
\end{equation}
where $q_\text{CH1}(q_\text{CH2})$ is the integrated charge of channel 1 (channel 2) over the full waveform, is close to zero for scintillation events that deposit light symmetrically into the two SiPMs. Second, the charge-weighted \textit{meantime} parameter,
\begin{equation}
    \text{Meantime} \equiv \frac{\int_{t_0} q\cdot t \,dt}{\int_{t_0}t\,dt},
\end{equation}
evaluated over a window from the trigger time $t_0$ (denoted Meantime$_{300}$ for integration from $t_0$ to $t_0 + 300$\,ns), separates prompt scintillation events from noise events. These parameters are illustrated in Fig.~\ref{fig:es}, which are in terms of energy.

\begin{figure}[t]
\centerline{\includegraphics[width=3.5in]{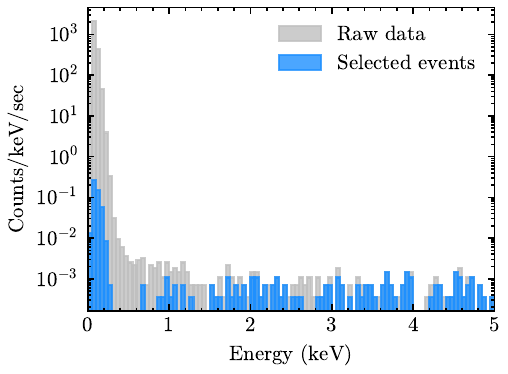}}
\caption{Low-energy spectrum of the 15.5\,h background run at 240\,K. Raw data is presented with grey bar and the spectrum after the simple event selections is presented with blue bar, reaching an analysis threshold around 0.35\,keV.}
\label{fig:espectrum}
\end{figure}

Utilizing the identical detector, a duration of 15.5\,h of background data was collected at a constant temperature of nearly 240\,K, within the temperature region of enhanced light yield. The energy spectrum before and after the event selection is shown in Fig.~\ref{fig:espectrum}. 
The combination of the two simple cuts effectively suppresses low-energy noise, achieving an analysis threshold of approximately 0.35 keV.

\section{Conclusion}
\label{sec:conclusion}
We have characterized a NaI(Tl) crystal readout by two directly coupled SiPMs in a dual-channel configuration over 94--293\,K. The internal and external optical crosstalk of the SiPMs were characterized and corrected, and the light yield rises from $13.2\pm0.8$\,PE/keV at room temperature and peaked $17.7\pm1.1$\,PE/keV at 238\,K, where the energy resolution reaches $5.3\pm0.2$\%. The scintillation decay shows the expected two-component temperature-dependence, with the slow component exhibiting a maximum near the temperature that has maximum light yield. 

This is the first dual-channel SiPM readout of a NaI(Tl) crystal characterized across this temperature range, demonstrating that cryogenic SiPM readout provides both enhanced scintillation performance and substantial noise suppression relative to conventional PMT-based systems.
Exploiting a two-step event selection based on charge asymmetry and meantime cuts, a background run at 240\,K reaches an analysis threshold around 0.35\,keV.
This improvement is the result of two complementary effects: the dual-channel readout allows noise triggering, and cryogenic operation reduces the DCR from SiPM by more than two orders of magnitude between room temperature and 240 K, with higher NaI(Tl) light yield.
The achieved threshold is half of the 0.7\,keV threshold reported for COSINE-100 experiment, which is PMT-based at room temperature and applying  machine-learning-based elaborated event selection method~\cite{yu2024}.

While the baseline design of the current COSINE-100U experiment still employs PMTs~\cite{lee2025}, this work demonstrates SiPMs as a complementary photosensor option for low-temperature NaI(Tl) operation. The resulting improvements in light yield and noise suppression make the SiPM--NaI(Tl) module a promising building block for extending the low-mass dark matter sensitivity of the COSINE experiment and its successor.

\bibliographystyle{IEEEtran}
\bibliography{sipmnai}

@article{schumann2019,
  title = {Direct {{Detection}} of {{WIMP Dark Matter}}: {{Concepts}} and {{Status}}},
  author = {Schumann, Marc},
  year = 2019,
  journal = {J. Phys. G},
  volume = {46},
  number = {10},
  pages = {103003},
  doi = {10.1088/1361-6471/ab2ea5},
}

@article{lightfoot,
title = {Characterisation of a silicon photomultiplier device for applications in liquid argon based neutrino
physics and dark matter searches},
author = {Lightfoot, P. K. and others},
year = 2008,
journal = {J. Instrum.},
volume = {3},
pages = {P10001},
doi = {10.1088/1748-0221/3/10/P10001}
}

@article{lee2025,
  title = {Upgrading the {{COSINE-100}} Experiment for Enhanced Sensitivity to Low-Mass Dark Matter Detection},
  collaboration = {COSINE-100},
  author = {Lee, Doohyeok and others},
  year = 2025,
  journal = {Commun. Phys.},
  volume = {8},
  number = {1},
  pages = {135},
  publisher = {Nature Publishing Group},
  issn = {2399-3650},
  doi = {10.1038/s42005-025-02067-4},
  urldate = {2026-01-17},
  copyright = {2025 The Author(s)},
}

@article{sailer2012,
  title = {Low Temperature Light Yield Measurements in {{NaI}} and {{NaI}}({{Tl}})},
  author = {Sailer, C. and Lubsandozhiev, B. and Strandhagen, C. and Jochum, J.},
  year = 2012,
  journal = {Eur. Phys. J. C},
  volume = {72},
  pages = {2061},
  doi = {10.1140/epjc/s10052-012-2061-7}
}

@article{lee2022a,
title = {{Study on NaI(Tl) Crystal at -35 °C for Dark Matter Detection}},
  author = {Lee, S. H. and Kim, G. S. and Kim, H. J. and Kim, K. W. and Lee, J. Y. and Lee, H. S.},
  year = 2022,
  journal = {Astropart. Phys.},
  volume = {141},
  pages = {102709},
  issn = {0927-6505},
  doi = {10.1016/j.astropartphys.2022.102709},
}

@article{park2026,
  title = {Validation of the COSINE-100U {{NaI}}({{Tl}}) encapsulation for low-temperature operation in liquid scintillator},
  author = {Park, K. H. and others},
   year = 2026,
   journal = {J. Instrum.},
   volume = {21},
   pages = {T03010},
   publisher = {IOP Publishing},
   doi = {10.1088/1748-0221/21/03/T03010}
}

@article{lee2022,
  title = {Scintillation Characteristics of a {{NaI}}({{Tl}}) Crystal at Low-Temperature with Silicon Photomultiplier},
  author = {Lee, H. Y. and Jeon, J. A. and Kim, K. W. and Kim, W. K. and Lee, H. S. and Lee, M. H.},
  year = 2022,

  journal = {J. Instrum.},
  volume = {17},
  number = {02},
  pages = {P02027},
  publisher = {IOP Publishing},
  issn = {1748-0221},
  doi = {10.1088/1748-0221/17/02/P02027},
}

@article{martinenghi2025,
  title = {First Use of Large Area {{SiPM}} Matrices Coupled with {{NaI}}({{Tl}}) Scintillating Crystal for Low Energy Dark Matter Search},
  collaboration = {ASTAROTH},
  author = {Martinenghi, Edoardo and others},
  year = 2025,

  journal = {Eur. Phys. J. C},
  volume = {85},
  number = {12},
  pages = {1444},
  issn = {1434-6052},
  doi = {10.1140/epjc/s10052-025-15197-4},
  urldate = {2026-01-14},
}

@article{adhikari2018b,
  title = {Initial {{Performance}} of the {{COSINE-100 Experiment}}},
  collaboration = {COSINE-100},
  author = {Adhikari, G. and others},
  year = 2018,
  journal = {Eur. Phys. J. C},
  volume = {78},
  number = {2},
  pages = {107},
  doi = {10.1140/epjc/s10052-018-5590-x},
}

@article{yu2024,
  title = {Lowering Threshold of {{NaI}}({{Tl}}) Scintillator to 0.7 {{keV}} in the {{COSINE-100}} Experiment},
  author = {Yu, G.H. and others},
  collaboration = {COSINE-100},
  year = 2024,

  journal = {J. Instrum.},
  volume = {19},
  number = {12},
  pages = {P12013},
  publisher = {IOP Publishing},
  issn = {1748-0221},
  doi = {10.1088/1748-0221/19/12/P12013},
  urldate = {2026-03-24},
}

@article{guan,
    title = {Study of Silicon Photomultiplier external cross-talk},
    author = {Guan, Y. and Anfimov, N. and Cao, G. and Xie, Z. and Dai, Q. and Fedoseev, D. and Kuznetsova, K. and Rybnikov, A. and Selyunin, A. and Sotnikov, A.},
    doi = {10.1088/1748-0221/19/06/P06024},
    journal = {J. Instrum.},
    volume = {19},
    pages = {P06024},
    year = {2024}
}

@article{otte2017,
  title = {Characterization of Three High Efficiency and Blue Sensitive Silicon Photomultipliers},
  author = {Otte, Adam Nepomuk and Garcia, Distefano and Nguyen, Thanh and Purushotham, Dhruv},
  year = 2017,

  journal = {Nucl. Instrum. Methods Phys. Res. A},
  volume = {846},
  pages = {106--125},
  issn = {0168-9002},
  doi = {10.1016/j.nima.2016.09.053},
  urldate = {2023-05-01},
}

@article{konovalov2024,
    title = {Status of the {COHERENT} experiment and new physics opportunities at the SNS Facility},
    author = {Konovalov, A. M},
    year = 2024,
    journal = {Moscow Univ. Phys. Bull.},
    volume = {79},
    pages = {220-226}
}

@article{COSINE:2020egt,
  title = {Development of Ultra-Pure {{NaI}}({{Tl}}) Detectors for the {{COSINE-200}} Experiment},
  author = {Park, B. J. and others},
  year = 2020,
  journal = {Eur. Phys. J. C },
  volume = {80},
  number = {9},
  pages = {814},
  doi = {10.1140/epjc/s10052-020-8386-8},
  collaboration = {COSINE}
}

@article{kim2025,
  title = {Scintillation Characteristics of an Undoped {{CsI}} Crystal at Low-Temperature for Dark Matter Search},
  author = {Kim,~W.~K. and others},
  year = 2025,

  journal = {Astropart. Phys.},
  volume = {173},
  pages = {103150},
  issn = {09276505},
  doi = {10.1016/j.astropartphys.2025.103150},
  urldate = {2026-01-30},
}

@article{Wang:2022ekc,
    author = "Wang, Lei and others",
    title = "{Reactor neutrino physics potentials of cryogenic pure-CsI crystal}",
    primaryClass = "physics.ins-det",
    doi = "10.1140/epjc/s10052-024-12800-y",
    journal = "Eur. Phys. J. C",
    volume = "84",
    number = "4",
    pages = "440",
    year = "2024"
}

@article{dietrich1973,
  title = {Kinetics of Self-Trapped Holes in Alkali-Halide Crystals: {{Experiments}} in {{NaI}}({{Tl}}) and {{KI}}({{Tl}})},
  author = {Dietrich, H.B. and Purdy, A.E. and Murray, R.B. and Williams, R.T.},
  year = 1973,
  journal = {Phys. Rev. B},
  volume = {8},
  number = {12},
  pages = {5894--5901},
  doi = {10.1103/PhysRevB.8.5894}
}

@article{adhikari2018,
  title = {An Experiment to Search for Dark-Matter Interactions Using Sodium Iodide Detectors},
  author = {Adhikari, Govinda and others},
  collaboration = {COSINE-100},
  year = 2018,
  journal = {Nature},
  volume = {564},
  number = {7734},
  pages = {83},
  doi = {10.1038/s41586-018-0739-1}
}

@article{robert1949,
    title = {The Detection of Gamma-Rays with Thallium-Activated Sodium Iodide Crystals},
    author = {Robert Hofstadter},
    year = {1949},
    journal = {Phys. Rev.},
    volume = {75},
    number = {1611},
    doi = {10.1103/PhysRev.75.796}
}

@article{antonello2019,
  title = {The {{SABRE}} Project and the {{SABRE Proof-of-Principle}}},
  author = {Antonello, M. and others},
  year = 2019,
  journal = {Eur. Phys. J. C},
  volume = {79},
  number = {4},
  pages = {363},
  doi = {10.1140/epjc/s10052-019-6860-y}
}

@article{amare2019b,
  title = {Performance of {{ANAIS-112}} Experiment after the First Year of Data Taking},
  author = {Amare, J. and others},
  collaboration = {ANAIS-112},
  year = 2019,
  journal = {Eur. Phys. J. C},
  volume = {79},
  number = {3},
  pages = {228},
  doi = {10.1140/epjc/s10052-019-6697-4}
}

@article{Bernabei:2020mon,
  title = {The {{DAMA}} Project: {{Achievements}}, Implications and Perspectives},
  author = {Bernabei, R. and others},
  year = 2020,
  journal = {Prog. Part. Nucl. Phys.},
  volume = {114},
  pages = {103810},
  doi = {10.1016/j.ppnp.2020.103810}
}

@article{carlin2025,
  title = {{{COSINE-100}} Full Dataset Challenges the Annual Modulation Signal of {{DAMA}}/{{LIBRA}}},
  author = {Seung Mok Lee and others},
  collaboration = {{COSINE-100}},
  year = 2025,
  journal = {Sci. Adv.},
  volume = {11},
  number = {36},
  pages = {eadv650},
  doi = {10.1126/sciadv.adv650},
}

@article{Billard:2021uyg,
    author = "Billard, Julien and others",
    title = "{Direct detection of dark matter{\textemdash}APPEC committee report*}",
    eprint = "2104.07634",
    archivePrefix = "arXiv",
    primaryClass = "hep-ex",
    doi = "10.1088/1361-6633/ac5754",
    journal = "Rept. Prog. Phys.",
    volume = "85",
    number = "5",
    pages = "056201",
    year = "2022"
}

@article{goodman1985a,
  title = {Detectability of {{Certain Dark Matter Candidates}}},
  author = {Goodman, Mark W. and Witten, Edward},
  year = 1985,
  journal = {Phys. Rev. D},
  volume = {31},
  pages = {3059},
  doi = {10.1103/PhysRevD.31.3059}
}

@article{ANFIMOV2021165162,
title = {Study of silicon photomultiplier performance at different temperatures},
journal = {Nucl. Instrum. Methods Phys. Res. A},
volume = {997},
pages = {165162},
year = {2021},
issn = {0168-9002},
doi = {https://doi.org/10.1016/j.nima.2021.165162},
author = {N. Anfimov and D. Fedoseev and A. Rybnikov and A. Selyunin and S. Sokolov and A. Sotnikov}
}

@article{DINU2015275,
title = {Studies of MPPC detectors down to cryogenic temperatures},
journal = {Nucl. Instrum. Methods Phys. Res. A},
volume = {787},
pages = {275-279},
year = {2015},
issn = {0168-9002},
doi = {https://doi.org/10.1016/j.nima.2014.12.061},
author = {N. Dinu and A. Nagai and A. Para}
}

\end{document}